\documentclass[aps,prl,twocolumn,nofootinbib, superscriptaddress,10pt]{revtex4-1}
\usepackage{graphicx}
\usepackage{epstopdf}
\usepackage{amsmath}
\usepackage{amssymb}
\usepackage{mathrsfs}
\usepackage{amsthm}
\usepackage{bm}
\usepackage{url}
\usepackage[T1]{fontenc}
\usepackage{csquotes}
\MakeOuterQuote{"}

\usepackage{color}

\newtheoremstyle{note}
  {\topsep/2}               % ABOVE SPACE
  {\topsep/2}               % BELOW SPACE
  {}                      % BODY FONT
  {\parindent}            % INDENT (empty value is the same as 0pt)
  {\itshape}              % HEAD FONT
  {.}                     % HEAD PUNCTUATION
  {5pt plus 1pt minus 1pt}% HEAD SPACE
  {}

\theoremstyle{note}
\newtheorem{theorem}{Theorem}
\newtheorem{lemma}{Lemma}

\theoremstyle{definition}

\theoremstyle{remark}

\newcommand{\tr}{\operatorname{tr}}

 \newcommand{\rme}{\mathrm{e}}

 \newcommand{\na}{\mathrm{NA}}

 \newcommand{\caH}{\mathcal{H}}

 \newcommand{\nni}{\mathbb{Z}^{\geq0}}

 \newcommand{\id}{1}

\newcommand{\be}{\begin{equation}}
\newcommand{\ee}{\end{equation}}
\newcommand{\ba}{\begin{align}}
\newcommand{\ea}{\end{align}}

\def\<{\langle}  %% overriding the original command \<
\def\>{\rangle}  %% overriding the original command \>

\def\eqref#1{\textup{(\ref{#1})}}  %% overriding the original command \eqref
\newcommand{\eref}[1]{Eq.~\textup{(\ref{#1})}}

\newcommand{\fref}[1]{Fig.~\ref{#1}}

\newcommand{\thref}[1]{Theorem~\ref{#1}}
\newcommand{\Thref}[1]{Theorem~\ref{#1}}

\newcommand{\cref}[1]{Conjecture~\ref{#1}}
\newcommand{\Cref}[1]{Conjecture~\ref{#1}}

\newcommand{\rcite}[1]{Ref.~\cite{#1}}
\newcommand{\rscite}[1]{Refs.~\cite{#1}}

\begin{document}
\title{Efficient Verification of Pure Quantum States in the Adversarial Scenario}

\author{Huangjun Zhu}
\email{zhuhuangjun@fudan.edu.cn}

\affiliation{Department of Physics and Center for Field Theory and Particle Physics, Fudan University, Shanghai 200433, China}

\affiliation{State Key Laboratory of Surface Physics, Fudan University, Shanghai 200433, China}

\affiliation{Institute for Nanoelectronic Devices and Quantum Computing, Fudan University, Shanghai 200433, China}

\affiliation{Collaborative Innovation Center of Advanced Microstructures, Nanjing 210093, China}

\author{Masahito Hayashi}
\affiliation{Graduate School of Mathematics, Nagoya University, Nagoya, 464-8602, Japan}

\affiliation{Shenzhen Institute for Quantum Science and Engineering,
Southern University of Science and Technology,
%No.1088 Xueyuan Avenue,Nanshan District,
Shenzhen,
518055, China}
\affiliation{Center for Quantum Computing, Peng Cheng Laboratory, Shenzhen 518000, China}
\affiliation{Centre for Quantum Technologies, National University of Singapore, 3 Science Drive 2, 117542, Singapore}

\begin{abstract}
Efficient verification of pure quantum states in the adversarial scenario is crucial to many applications in quantum information processing,  such as blind measurement-based quantum computation and quantum networks. However, little is known about this topic so far.  Here
we establish a general framework for verifying pure quantum states in the adversarial scenario and clarify the resource cost. 
Moreover,   we propose a simple and general recipe to constructing efficient verification protocols for the adversarial scenario from  protocols for the nonadversarial scenario. With this recipe, arbitrary pure states can be verified in the adversarial scenario with almost the same efficiency as in the nonadversarial scenario. Many important quantum states  can be verified  in the adversarial scenario using  local projective measurements with unprecedented high efficiencies. 
\end{abstract}

\date{\today}
\maketitle

\emph{Introduction.}---Bipartite and multipartite entangled states play a central role in quantum information processing and foundational studies \cite{HoroHHH09,GuhnT09}. Accurate preparation and verification of desired quantum states is a key to various applications. However,  characterization methods based on traditional tomography are  inefficient as the resource required  grows exponentially with the number of qubits.  
Even popular alternatives, such as compressed sensing \cite{GrosLFB10} and direct fidelity estimation (DFE) \cite{FlamL11}, cannot avoid this scaling behavior. Recently, a powerful approach known as quantum state verification (QSV) \cite{HayaMT06,Haya09G,PallLM18} has attracted increasing attention. Efficient protocols based on local measurements have been constructed for  bipartite pure states \cite{HayaMT06,Haya09G,PallLM18,ZhuH19O,LiHZ19,WangH19,YuSG19}, stabilizer states (including graph states) \cite{HayaM15,FujiH17,HayaH18, PallLM18,MarkK18}, hypergraph states \cite{ZhuH19E}, weighted graph states \cite{HayaT19}, and Dicke states~\cite{LiuYSZ19}.

The problem is much more complicated in  the adversarial scenario, in which the states to be verified are prepared by a  malicious  adversary. Efficient 
verification of quantum states in this scenario is a key
to many important  applications, such as blind measurement-based quantum computation (MBQC) \cite{MoriF13,HayaM15,FujiH17,HayaH18,TakeMH19} and quantum networks \cite{PersLCL13,MccuPBM16,MarkK18}. However, little is known on this topic. The  approach proposed in \rcite{PallLM18} does not apply although it is quite successful in the nonadversarial scenario. Other approaches known in the literature only apply to certain special types of states and are highly inefficient. 
To verify hypergraph states with recent approaches in \rscite{MoriTH17,TakeM18} for example, the number of required tests is 
enormous even in the simplest nontrivial cases. An outstanding problem underlying this deadlock is that, even for a given verification strategy, no efficient method is known for determining the minimal number of tests required to achieve a given precision, as characterized by the infidelity and significance level \cite{TakeM18,TakeMMM19}.

In this paper 
we establish a general framework of QSV in  the adversarial scenario and settle several fundamental problems. First, we determine the precision achievable with  a given strategy and a given  number of tests and thereby clarify the resource cost to achieve a given precision. Then we propose a general recipe to constructing efficient verification protocols for the adversarial scenario from  protocols for the nonadversarial scenario. With this recipe, arbitrary pure states can be verified in the adversarial scenario with almost the same efficiency as in the nonadversarial scenario. For high-precision verification, the overhead in the number of tests is at most three times. Together with recent works, this recipe
can be applied immediately to  efficiently verify  many important quantum states, such as  bipartite pure states, stabilizer states (including graph states), hypergraph states, weighted graph states, and Dicke states, even if we can only perform local projective measurements.

This paper extracts the key results in \rcite{ZhuH19AdL}, which contains complete technical details and additional results, including the proofs of all statements presented here.

\emph{Verification of a pure state.}---
Consider a device that is supposed to  produce some target state $|\Psi\>$ in the  Hilbert space ${\cal H}$, but actually  produces $\sigma_1, \sigma_2, \ldots, \sigma_{N}$ in $N$ runs. Our task is to verify whether each $\sigma_j$ is sufficiently close to the target state on average. To achieve this task we can perform two-outcome  measurements $\{E_l, 1-E_l\}$ randomly from a set of accessible measurements in each run. Each  measurement is specified by a test operator $E_l$, which corresponds to passing the test and satisfies $E_l|\Psi\>=|\Psi\>$, so that the target state can alway pass the test. After $N$ runs, we accept the source if and only if it passes all tests. 
 Suppose the test $E_l$ is performed with probability $\mu_l$; then the efficiency of the verification strategy is determined by the verification operator $\Omega:=\sum_l \mu_l E_l$. If $\<\Psi|\sigma_j|\Psi\>\leq 1-\epsilon$; then 
the maximal probability that $\sigma_j$ can pass each test on average
reads 
\begin{equation}\label{eq:PassingProb}
\max_{\<\Psi|\sigma|\Psi\>\leq 1-\epsilon }\tr(\Omega \sigma)=1- [1-\beta(\Omega)]\epsilon=1- \nu(\Omega)\epsilon, 
\end{equation}
where the maximization is taken over all quantum states $\sigma$ that satisfy $\<\Psi|\sigma|\Psi\>\leq 1-\epsilon$ \cite{PallLM18,ZhuH19AdL}. 
Here  $\beta(\Omega)$ is the second largest eigenvalue of $\Omega$, and  $\nu(\Omega):=1-\beta(\Omega)$ is the spectral gap from the maximal eigenvalue.

Suppose the outputs $\sigma_1, \sigma_2, \ldots, \sigma_N$ of the device are independent of each other. Let $\epsilon_j=1-\<\Psi|\sigma_j |\Psi\>$ be the infidelity between $\sigma_j$ and $|\Psi\>$. Then these states can pass $N$ tests with probability at most
\begin{align}\label{eq:PassingProbN}
\prod_{j=1}^N \tr(\Omega\sigma_j)\leq \prod_{j=1}^N [1-\nu(\Omega)\epsilon_j]\leq [1-\nu(\Omega)\bar{\epsilon}]^N, 
\end{align}
where $\bar{\epsilon}=\sum_j \epsilon_j/N$ is the average infidelity.
The bound in \eref{eq:PassingProbN} is saturated when all $\epsilon_j$ are equal and each $\sigma_j$
 is supported in the subspace of $\caH$  associated with the two largest eigenvalues of  $\Omega$. So passing $N$ tests can  ensure the condition
$\bar{\epsilon}<\epsilon$ with significance level $\delta= [1-\nu(\Omega)\epsilon]^N$,  where the significance level is the maximal probability of accepting the source when $\bar{\epsilon}\geq \epsilon$. Accordingly, to verify $|\Psi\>$ within given infidelity $\epsilon$ and  significance level $\delta$,  the minimum number of tests can be determined by minimizing the number $N$ that satisfies the inequality $[1-\nu(\Omega)\epsilon]^N\leq \delta$, with the result
\begin{equation}\label{eq:NumTest}
\!N_\na(\epsilon,\delta,\Omega)\!=\!\biggl\lceil
\frac{1}{\ln[1-\nu(\Omega)\epsilon]}\ln\delta\biggr\rceil
\leq 
\biggl\lceil
\frac{1}{\nu(\Omega)\epsilon}\ln\frac{1}{\delta}\biggr\rceil.
\end{equation}

A  formula similar to \eref{eq:NumTest} was previously derived in \rcite{PallLM18};  however, here the underlying assumption and the interpretation are quite different. Notably, we do not require the unnatural assumption that either $\<\Psi|\sigma_j|\Psi\>= 1$ for all $j$ or  $\<\Psi|\sigma_j|\Psi\>\leq 1-\epsilon$ for all $j$. In addition, our conclusion 
concerns the average fidelity, which is more relevant than the  maximal fidelity addressed in \rcite{PallLM18}. It should be pointed out that the above conclusion is meaningful only if the states produced 
after the verification procedure have the same average fidelity as  in the verification. This assumption is reasonable and is usually taken for granted in experiments if the source is not malicious. If this assumption fails, then we have to apply a protocol tailored for the adversarial scenario to be discussed shortly.

In view of \eref{eq:NumTest}, to minimize the number of tests, we need to maximize the spectral gap. If there is no restriction on the accessible measurements, then the optimal strategy consists of the single test $\{|\Psi\>\<\Psi|, 1-|\Psi\>\<\Psi|\}$, so that  $\Omega=|\Psi\>\<\Psi|$, $\beta(\Omega)=0$, and $\nu(\Omega)=1$; cf.~\rcite{PallLM18}. In practice, we need to consider various constraints on measurements. In addition,  the situation for the adversarial scenario is quite different as we shall see.

\emph{Adversarial scenario.}---In the adversarial scenario, the device is controlled by a potentially malicious adversary and can produce an arbitrarily correlated or even entangled state $\rho$  on $\caH^{\otimes (N+1)}$, as encountered in blind MBQC.
For example, the device can prepare  $(|\Psi\>\<\Psi|)^{\otimes (N+1)}$ with probability $0<a<1$ and $\sigma^{\otimes (N+1)}$ with probability $1-a$.  In this case, the above approach and the variant in \rcite{PallLM18} are not applicable  as analyzed in   \rcite{ZhuH19AdL}.
It turns out they can be applied to the adversarial scenario after some modification, but the analysis on the minimal number of tests will be completely different. Here we shall clarify this issue and propose a simple and efficient recipe to QSV in the adversarial scenario.

 To verify the state produced, we randomly choose $N$ systems and apply a certain  strategy $\Omega$  to each   system chosen.  Our goal is to ensure that the reduced state on the remaining system has fidelity at least $1-\epsilon$ if $N$ tests are passed.  Since $N$ systems are chosen randomly, without loss of generality, we  may assume that $\rho$ is permutation invariant.
Suppose the strategy  $\Omega$ is applied to the first $N$ systems, then  the probability that $\rho$ can   pass $N$ tests reads $p_\rho=\tr[(\Omega^{\otimes N}\otimes \id) \rho]$. 
The reduced state  on system $N+1$ (assuming $p_\rho>0$) is given by
$\sigma'_{N+1}=p_\rho^{-1}\tr_{1,2,\ldots, N}[(\Omega^{\otimes N}\otimes \id) \rho],
$
where $\tr_{1,2,\ldots, N}$ means the partial trace over the systems $1,2,\ldots, N$. The fidelity between $\sigma'_{N+1}$ and $|\Psi\>$  reads $F_\rho=\<\Psi|\sigma'_{N+1}|\Psi\>=p_\rho^{-1} f_\rho$, 
where $
f_\rho=\tr[(\Omega^{\otimes N}\otimes |\Psi\>\<\Psi|) \rho]$.

To  characterize the performance of the strategy $\Omega$ adapted to the adversarial scenario,  define
\begin{align}
F(N,\delta,\Omega)
&:=\min_{\rho}
\big\{
p_\rho^{-1}f_\rho
\,|\,p_\rho \ge \delta
\big\}, \quad  0<\delta\leq 1.  \label{eq:MinFdelta}
\end{align}
This figure of merit  denotes the minimum fidelity of $\sigma'_{N+1}$ with the target state suppose that $\rho$  can pass $N$ tests with probability
at least $\delta$; it is
 nondecreasing in $\delta$ by definition.
For $0<\epsilon,\delta< 1$,  define $N(\epsilon,\delta,\Omega)$ as the minimum number of tests required to verify $|\Psi\>$ within infidelity $\epsilon$ and significance level $\delta$, that is, 
\begin{equation}\label{eq:MinNumTestDef}
N(\epsilon,\delta,\Omega):=\min\{N\geq 1\,| \, F(N,\delta,\Omega)\geq 1-\epsilon \}. 
\end{equation}

\emph{Homogeneous strategies.}---A strategy (or verification operator) $\Omega$ for $|\Psi\>$ is \emph{homogeneous} if it has the form 
\begin{equation}\label{eq:HomoStrategy}
\Omega=|\Psi\>\<\Psi|+\lambda(\id-|\Psi\>\<\Psi|),
\end{equation}
where $0\leq \lambda<1$. In this case, all eigenvalues of $\Omega$ are equal to $\lambda$ except for the largest one, so  we have $\beta=\lambda$ and $\nu=1-\lambda$. Now  it is natural and more informative to 
replace $\Omega$ with $\lambda$ in the notations of various figures of merit; for example, we can write $F(N,\delta,\lambda)$ in place of  $F(N,\delta,\Omega)$. A homogeneous strategy is the most efficient among all verification strategies with  a given spectral gap and so plays a key role in QSV.

When $\lambda=0$, the verification operator $\Omega$  is singular (has a zero eigenvalue). For $0<\delta\leq1$, calculation shows that
\begin{align}
F(N,\delta,\lambda=0)=\max\biggl\{0,\;
\frac{(N+1)\delta-1}{N\delta } \biggr\}. 
\label{eq:FidelityHomo0}
\end{align}
To verify the  target state  within infidelity $\epsilon$ and significance level $\delta$, the minimum number  of tests required is
\begin{equation}\label{eq:NumTestSingHomo}
N(\epsilon,\delta,\lambda=0)=\biggl \lceil\frac{1-\delta}{\epsilon\delta}\biggr\rceil.
\end{equation}
The scaling with $1/\delta$ is suboptimal although the strategy is optimal for the nonadversarial scenario by \eref{eq:NumTest} when there is no restriction on the accessible measurements.

\begin{figure}
	\includegraphics[width=8cm]{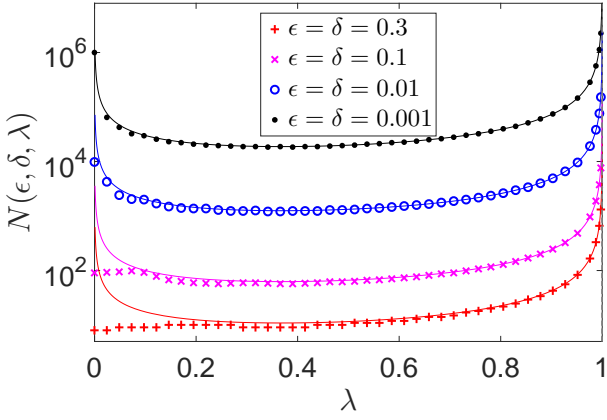}
	\caption{\label{fig:NumTest}(color online)
Number of tests $N(\epsilon,\delta,\lambda)$ required to verify a pure  state within infidelity $\epsilon$ and significance level $\delta$ in the adversarial scenario using a homogeneous strategy characterized by $\lambda$. 
For comparison, the  approximate formula $(\ln\delta)/(\lambda\epsilon\ln\lambda)$ is plotted as curves.	
}
\end{figure}

When  $0<\lambda<1$, the verification operator $\Omega$  is nonsingular (positive definite). Let  $\nni$ denote the set of nonnegative integers. For $k\in \nni$, define
\begin{align}
\zeta(N,\delta,\lambda,k)&:= 
\frac{\lambda\{\delta[1+(N-k)\nu]-\lambda^k\}}{\nu(k\nu+N\lambda)}. 
 \label{eq:zetadel}
\end{align}
The following theorem clarifies the precision that can be achieved by a homogeneous strategy given $N$ tests.
\begin{theorem}\label{thm:FidelityHomo}
	Suppose $0<\lambda <1$; then $F(N,\delta,\lambda)=0$ if $0<\delta \leq \lambda^N$ and $F(N,\delta,\lambda)=\zeta(N,\delta,\lambda,k_*)/\delta$ if instead $\lambda^N<\delta\leq 1$,
where $k_*$ is the largest integer $k$ that satisfies $(N+1-k)\lambda^k+k\lambda^{k-1}\geq (N+1)\delta$. 
\end{theorem}
Let $k_+:=\lceil\log_\lambda\delta\rceil$ and  $k_-:=\lfloor\log_\lambda\delta\rfloor$; then 
$k_*$  equals  either $k_+$ or $k_-$ given the assumption $\lambda^N<\delta\leq 1$. Define
\begin{align}
\tilde{N}(\epsilon,\delta,\lambda,k)&:=\frac{k\nu^2 \delta F +\lambda^{k+1}+\lambda\delta(k\nu-1)}{\lambda\nu\delta \epsilon}, \label{eq:NumTestHomoC}
\end{align}
where $F=1-\epsilon$ and $\nu=1-\lambda$. The following two theorems provide analytical formulas and informative bounds for $N(\epsilon,\delta,\lambda)$.  The results are illustrated in \fref{fig:NumTest}.
\begin{theorem}\label{thm:NumTestHomo}
	Suppose $0<\epsilon,\delta, \lambda<1$. Then 
	\begin{align}
	N(\epsilon, \delta,\lambda)&=\Bigl\lceil\min_{k\in \nni}\!\tilde{N}(\epsilon,\delta,\lambda,k)\Bigr\rceil=\bigl\lceil\tilde{N}(\epsilon,\delta,\lambda,k^*)\bigr\rceil, \label{eq:NumTestHomo1}
	\end{align}
where $k^*$ is the largest integer $k$ that obeys the inequality $\delta\leq \lambda^{k}/(F\nu+\lambda )= \lambda^{k}/(F+\lambda \epsilon)$. 
\end{theorem}
\begin{theorem}\label{thm:NumTestHomoBounds} Suppose $0<\epsilon,\delta,\lambda <1$. Then 
	\begin{gather}
k_-+\biggl\lceil\frac{k_-F}{\lambda\epsilon}\biggr\rceil\leq N(\epsilon,\delta,\lambda)\leq	 \biggl\lceil\frac{\ln\delta}{\lambda\epsilon\ln \lambda}-\frac{\nu k_-}{\lambda}\biggr\rceil .\label{eq:NumberBoundAdvHomo}
\end{gather}	
Both the upper bound and lower bound  are saturated when $(\ln\delta)/\ln\lambda$ is an integer. 
\end{theorem}

In the high-precision limit $\epsilon,\delta\rightarrow 0$,  $k_\pm\approx(\ln\delta)/\ln\lambda$, so  \thref{thm:NumTestHomoBounds} implies that
\begin{equation}\label{eq:NumTestHomoApp}
N(\epsilon,\delta,\lambda)\approx
(\lambda\epsilon\ln\lambda)^{-1}
\ln\delta=(\lambda\epsilon\ln\lambda^{-1})^{-1}\ln\delta^{-1}.
\end{equation}
The efficiency  is characterized by the factor $(\lambda\ln\lambda^{-1})^{-1}$, as reflected in \fref{fig:NumTest}. 
The number of tests is minimized when $\lambda=1/\rme$ (with $\rme$ being the base of the natural logarithm),  in which case  $N(\epsilon,\delta,\lambda=\rme^{-1})
\approx\rme\epsilon^{-1}
\ln\delta^{-1}$, which  is comparable to the counterpart $\epsilon^{-1}
\ln\delta^{-1}$ for the nonadversarial scenario.

\emph{General verification strategies.}---Now we turn to a general verification strategy $\Omega$; let $\beta=\beta(\Omega)$ and $\nu=\nu(\Omega)$.  
\begin{theorem}\label{thm:NIID}
Suppose $0<\delta,\nu\leq 1$. Then
	\begin{align}
	F(N,\delta,\Omega) \ge
	1-\frac{1- \delta}{ N\nu\delta};
	\label{eq:MinFidelityBound}
	\end{align}
the inequality  is saturated for $\frac{1+ N\beta}{N+1}\leq \delta \leq 1 $.
\end{theorem}
\Thref{thm:NIID} implies that 
\begin{equation}\label{eq:NumTestAdvUB}
N(\epsilon,\delta,\Omega)\leq\biggl\lceil
\frac{1- \delta}{\nu\delta \epsilon}\biggr\rceil.
\end{equation}
This bound   is much smaller than previous results based on the quantum de Finetti theorem \cite{MoriTH17,TakeM18}. Nevertheless, the scaling with $1/\delta$ is suboptimal, and this behavior is inevitable  if $\Omega$ is singular; cf.~\eref{eq:NumTestSingHomo}.

For a nonsingular verification operator $\Omega$,  the efficiency is mainly determined by its  second largest eigenvalue $\beta$ (or  $\nu=1-\beta$) and the smallest eigenvalue  $\tau$. Let $\tilde{\beta}:=
\beta$ if $\beta\ln \beta^{-1}\leq \tau\ln \tau^{-1}$
and $\tilde{\beta}:=
\tau$ otherwise.
\begin{lemma}\label{lem:Fboundf}
	Suppose $0<\delta\leq 1$, and  $\Omega$ is a nonsingular verification operator. Then
\begin{align}
F(N,\delta,\Omega)&\geq \frac{N+1-(\ln \beta)^{-1}\ln (\tau\delta)}{N+1-(\ln \beta)^{-1}\ln (\tau\delta)-h\ln (\tau\delta)},  \label{eq:Fbounddel}
\end{align}
where $h=(\tilde{\beta}\ln \tilde{\beta}^{-1})^{-1}=[\min\{\beta \ln \beta^{-1}, \tau \ln\tau^{-1}\}]^{-1}$. 
\end{lemma}

\begin{theorem}\label{thm:NumTestBounds}
	Suppose $0<\epsilon,\delta<1$, and  $\Omega$ is a nonsingular verification operator. Then
\begin{gather}
		k_-(\tilde{\beta})+\biggl\lceil\frac{k_-(\tilde{\beta})F}{\tilde{\beta}\epsilon}\biggr\rceil\leq 
		N(\epsilon,\delta,\Omega)< \frac{h\ln(F\delta)^{-1}}{\epsilon}, 	\label{eq:NumTestLUB}
\end{gather}
	where $F=1-\epsilon$ and   $k_-(\tilde{\beta})=\lfloor (\ln\delta)/\ln\tilde{\beta}\rfloor$. 
\end{theorem} 
In the limit $\epsilon,\delta\rightarrow 0$,
the upper and lower bounds in \eref{eq:NumTestLUB} are tight with respect to the relative deviation, so we have 
\begin{equation}\label{eq:NumberMeasOpt2}
N(\epsilon,\delta,\Omega)\approx  \frac{h\ln(\delta^{-1})}{\epsilon}=\frac{\ln\delta}{\epsilon\tilde{\beta}\ln\tilde{\beta}}.
\end{equation}
This number  has the same scaling behaviors with $\epsilon^{-1}$ and $\delta^{-1}$ as the counterpart for  the nonadversarial scenario  in \eref{eq:NumTest}. The  overhead is quantified by $\nu h=\nu/(\tilde{\beta}\ln\tilde{\beta}^{-1})$.

\emph{Recipe to constructing efficient protocols for the adversarial scenario.}---
The overhead of QSV in  the adversarial scenario could be quite large if the verification operator $\Omega$ is singular  or nearly singular. To resolve this problem  here  we introduce a general recipe  by adding the trivial test, where the  "trivial test"  means the test operator $E$  coincides with the identity operator, that is, $E=1$, so that  all quantum states can always pass the test.

Given a  verification operator $\Omega$ for the pure state $|\Psi\>$, we can  construct a \emph{hedged verification operator} as follows,
\begin{equation}
\Omega_p=p+(1-p)\Omega,\quad 0\le p<1. 
\end{equation}
It is realized by performing  
the trivial test and $\Omega$ with probabilities $p$ and $1-p$, respectively. The second largest and smallest  eigenvalues of $\Omega_p$ read
\begin{equation}
\beta_p=p+(1-p)\beta,\quad \tau_p=p+(1-p)\tau,
\end{equation}
where $\beta$ and $\tau$ are the second largest and  smallest eigenvalues  of $\Omega$, respectively. 
By \eref{eq:NumTestLUB}, to verify $|\Psi\>$ within infidelity $\epsilon$ and significance level $\delta$, the number of tests required by the strategy $\Omega_p$ (assuming $\tau_p>0$) satisfies
\begin{equation}\label{eq:NumTestAdvRev}
N(\epsilon,\delta,\Omega_p)< h(p,\nu,\tau)\epsilon^{-1} \ln(F\delta)^{-1}, 
\end{equation}
where $F=1-\epsilon$ and 
\begin{align}
h(p,\nu,\tau)&=\bigl[\min\bigl\{\beta_p \ln \beta_p^{-1}, \tau_p \ln \tau_p^{-1}\bigr\}\bigr]^{-1}. 
\end{align}	
Compared with  the nonadversarial scenario, the overhead satisfies
\begin{equation}\label{eq:AdvNaRatio1}
\frac{N(\epsilon,\delta,\Omega_{p})}{N_{\na}(\epsilon,\delta,\Omega)}<
\nu h(p,\nu,\tau)	\frac{[\ln (1-\nu\epsilon)^{-1}] \ln(F\delta)}{\nu\epsilon\ln \delta}. 
\end{equation}
This bound decreases monotonically with  $1/\epsilon$, $1/\delta$, and $1/\nu$ \cite{ZhuH19AdL}; it approaches $\nu h(p,\nu,\tau)$
in the limit $\epsilon, \delta\rightarrow 0$, in which case the bound is saturated.
So   the  function $\nu h(p,\nu,\tau)$
is of key interest to  characterizing the overhead of high-precision QSV in the adversarial scenario.

To achieve a high performance, we need to  minimize $h(p,\nu,\tau)$ over $p$.  The optimal probability $p$ reads
\begin{equation}
p_*(\nu,\tau)=\min\bigl\{p\geq 0| \beta_p\geq \rme^{-1}\;\& \;\tau_p \ln \tau_p^{-1}\geq \beta_p \ln \beta_p^{-1} \bigr\}, \label{eq:pstardef}
\end{equation}
which is nondecreasing in $\nu$ and nonincreasing in~$\tau$. 
For a  homogeneous strategy  $\Omega$ with $\tau=\beta=1-\nu$, we have $p_*(\nu,1-\nu)=(\rme \nu-\rme+1)/(\rme \nu)$ if $\nu\geq 1-(1/\rme)$ and  $p_*(\nu,1-\nu)=
0$ otherwise. When $\tau=0$, the probability $p_*(\nu):=p_*(\nu,0)$ can be approximated by $\nu/\rme$. In general,  it is  easy to compute $p_*(\nu,\tau)$ numerically.

\begin{figure}
	\includegraphics[width=6.1cm]{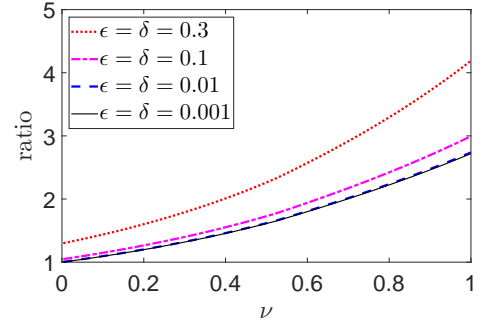}
	\caption{\label{fig:ratio}(color online)
	Overhead of QSV in the adversarial scenario compared with the nonadversarial scenario. Each curve represents an upper bound for the ratio $N(\epsilon,\delta,\Omega_{p})/N_{\na}(\epsilon,\delta,\Omega)$, where
$p=\nu/\rme$ and $\nu$ is the spectral gap of $\Omega$. The same bound holds if $p_*(\nu,\tau)\leq p\leq p_*(\nu)$; cf.~\eref{eq:AdvNaRatio}. 
		}
\end{figure}

\begin{theorem}\label{thm:AdvOHGen}If  $p=\nu/\rme$ or $p_*(\nu,\tau)\leq p\leq p_*(\nu)$, then 
	\begin{align}
	&N(\epsilon,\delta,\Omega_p)\!<\!  \frac{h(\nu/\rme,\nu,0) \ln(F\delta)^{-1}}{\epsilon}
\!\leq \! \frac{\ln(F\delta)^{-1}}{(1-\nu+\rme^{-1}\nu^2)\nu\epsilon}.\label{eq:NumTestGenA}
	\end{align}			
\end{theorem}
Here the number of tests  $N(\epsilon,\delta,\Omega_{p})$  achieves the optimal scaling behaviors in both
$\epsilon$ and $\delta$ as in the nonadversarial scenario, which have never been achieved before.
\Thref{thm:AdvOHGen} sets a general upper bound  on the overhead of QSV in the adversarial scenario. If $p=\nu/\rme$ or $p_*(\nu,\tau)\leq p\leq p_*(\nu)$ for example, then 
\begin{align}\label{eq:AdvNaRatio}
&\frac{N(\epsilon,\delta,\Omega_{p})}{N_{\na}(\epsilon,\delta,\Omega)}<
\nu h(\nu/\rme,\nu,0)	\frac{[\ln (1-\nu\epsilon)^{-1}] \ln(F\delta)}{\nu\epsilon\ln \delta}. 
\end{align}
Analysis shows that $h(\nu/\rme,\nu,0)$ decreases monotonically in $\nu$ (for $0<\nu\leq 1$), while
 $\nu h(\nu/\rme,\nu,0)$ increases monotonically and satisfies  $1<\nu h(\nu/\rme,\nu,0)\leq \rme$~\cite{ZhuH19AdL}. So the   bound in \eref{eq:AdvNaRatio} decreases monotonically with  $1/\epsilon$, $1/\delta$, and $1/\nu$, as illustrated in \fref{fig:ratio}.
The overhead is at most  three times when $\epsilon,\delta\leq 1/10$ and is  negligible as $\nu,\epsilon,\delta$ approach zero. Surprisingly,   we can choose the probability $p$ for performing the trivial test  without even knowing the value of $\tau$, while achieving a nearly optimal performance. In particular,  the choices $p=p_*(\nu)$ and $p=\nu/\rme$ are  nearly optimal. Meanwhile,  the performance of $\Omega_{p_*}$ is not sensitive to  $\tau$, unlike $\Omega$.

Furthermore, our recipe for the adversarial scenario requires the same measurement settings as required for the nonadversarial scenario except for the trivial test. So pure states can be verified in the adversarial scenario 
with almost the same efficiency as in the nonadversarial scenario with respect to both the total number of tests and the number of measurement settings. This conclusion holds even if we can only perform local measurements.

\emph{Summary.}---We established a general framework for verifying pure quantum states in  the adversarial scenario and clarified the resource cost. 
Moreover,  we proposed a simple but 
powerful recipe to constructing efficient verification protocols for the adversarial scenario from the counterpart for the  nonadversarial scenario. Thanks to this recipe, any pure state can be verified with almost the same efficiency as in the nonadversarial scenario. To construct an efficient protocol for the adversarial  scenario, it suffices to find an efficient protocol for the nonadversarial scenario  and then apply our recipe.

Our study can readily be applied to verifying many important quantum states in the adversarial scenario  with unprecedented high
efficiencies. In conjunction with recent works, optimal protocols based on local projective measurements can be constructed for all bipartite pure states which  require only $\lceil\rme \epsilon^{-1}\ln \delta^{-1}\rceil$ tests to achieve infidelity $\epsilon$ and significance level $\delta$. Nearly optimal protocols can be constructed for stabilizer states (including graph states) which require  $\lceil 3 \epsilon^{-1}\ln \delta^{-1}\rceil$ tests. General hypergraph states, weighted graph states, and Dicke states  can also be verified efficiently with about 
$n \epsilon^{-1}\ln \delta^{-1}$ tests, where 
$n$ is   the number of qubits. More details  can be found in the companion paper~\cite{ZhuH19AdL}. These results are instrumental to many applications in quantum information processing.

\acknowledgments
This work is  supported by  the National Natural Science Foundation of China (Grant No. 11875110). MH is supported in part by
 Fund for the Promotion of Joint International Research (Fostering Joint International Research) Grant No. 15KK0007,
 Japan Society for the Promotion of Science (JSPS) Grant-in-Aid for Scientific Research (A) No. 17H01280, (B) No. 16KT0017, and Kayamori Foundation of Informational Science Advancement.

%%%%%%%%%%%%%%%%%%%%%%%%%%%%%%%%%%%%%%%%%%%%%%%%%%%%%%%%%%%%%%%%%%%%%%%%%%%%%%%%%%%%%%%%

%@CONTROL{REVTEX41Control}
%@CONTROL{apsrev41Control,author="48",editor="1",pages="1",title="0",year="0"}

\nocite{apsrev41Control}
\bibliographystyle{apsrev4-1}
\bibliography{all_references}

\end{document}